\documentclass[prb,twocolumn,superscriptaddress,showkeys]{revtex4-1}

\usepackage{graphicx}
\usepackage[colorlinks=true,linkcolor=blue,urlcolor=blue,citecolor=blue]{hyperref}

\begin{document}

\title{Boosting local field enhancement by on-chip nanofocusing and impedance-matched plasmonic antennas}

\author{Vladimir A. Zenin}
\email{zenin@iti.sdu.dk}
\affiliation{Centre for Nano Optics, University of Southern Denmark, Campusvej 55, DK-5230 Odense M, Denmark}

\author{Andrei Andryieuski}
\affiliation{DTU Fotonik, Technical University of Denmark, Oersteds pl. 343, DK-2800 Kongens Lyngby, Denmark}

\author{Radu Malureanu}
\affiliation{DTU Fotonik, Technical University of Denmark, Oersteds pl. 343, DK-2800 Kongens Lyngby, Denmark}

\author{Ilya P. Radko}
\affiliation{Centre for Nano Optics, University of Southern Denmark, Campusvej 55, DK-5230 Odense M, Denmark}

\author{Valentyn~S.~Volkov}
\affiliation{Centre for Nano Optics, University of Southern Denmark, Campusvej 55, DK-5230 Odense M, Denmark}

\author{Dmitri K. Gramotnev}
\affiliation{Nanophotonics Pty. Ltd., GPO Box 786, Albany Creek, Queensland 4035, Australia}

\author{Andrei V. Lavrinenko}
\affiliation{DTU Fotonik, Technical University of Denmark, Oersteds pl. 343, DK-2800 Kongens Lyngby, Denmark}

\author{Sergey I. Bozhevolnyi}
\affiliation{Centre for Nano Optics, University of Southern Denmark, Campusvej 55, DK-5230 Odense M, Denmark}

\date{\today}


\begin{abstract}
Strongly confined surface plasmon-polariton modes can be used for efficiently delivering the electromagnetic energy to nano-sized volumes by reducing the cross sections of propagating modes far beyond the diffraction limit, i.e., by nanofocusing. This process results in significant local-field enhancement that can advantageously be exploited in modern optical nanotechnologies, including signal processing, biochemical sensing, imaging and spectroscopy. Here, we propose, analyze, and experimentally demonstrate on-chip nanofocusing followed by impedance-matched nanowire antenna excitation in the end-fire geometry at telecom wavelengths. Numerical and experimental evidences of the efficient excitation of dipole and quadrupole (dark) antenna modes are provided, revealing underlying physical mechanisms and analogies with the operation of plane-wave Fabry-P\'erot interferometers. The unique combination of efficient nanofocusing and nanoantenna resonant excitation realized in our experiments offers a major boost to the field intensity enhancement up to $\sim 12000$, with the enhanced field being evenly distributed over the gap volume of $30\times 30\times 10\ {\rm nm}^3$, and promises thereby a variety of useful on-chip functionalities within sensing, nonlinear spectroscopy and signal processing.
[This document is the unedited Author's version of a Submitted Work that was subsequently accepted for publication in \textit{Nano Letters}, \copyright American Chemical Society after peer review. To access the final edited and published work see \url{http://dx.doi.org/10.1021/acs.nanolett.5b03593}.]
\end{abstract}

\maketitle

The major aspect of focusing of electromagnetic radiation is the possibility of concentrating the energy in a small volume. Because of diffraction, the focusing of free-propagating optical waves is limited in size to the half of the light wavelength in the medium – the diffraction limit of light. \cite{ref1} One approach to overcome this limit is to use surface plasmon polaritons (SPPs) – surface electromagnetic modes bound to and propagating along metal-dielectric interfaces, with electromagnetic fields in a dielectric being coupled to collective free electron oscillations in a metal. \cite{ref2} Spatial confinement of SPP modes in the cross section perpendicular to the propagation direction depends on the material composition and geometric configuration of a waveguiding structure. Notably, some SPP modes (supported, for example, by metal nanowires \cite{ref3}) exhibit a unique scaling property in their spatial confinement: the mode is progressively better confined for smaller lateral waveguide dimensions, opening thereby the possibility for guiding extremely confined (i.e., on a deep subwavelength scale) SPP modes\cite{ref4} as well as for designing SPP-based nanoantennas. \cite{ref5} This feature can further be used for nanofocusing, which is the process of reducing the cross sections of propagating optical modes far beyond the diffraction limit, simply by gradually decreasing lateral waveguide dimensions. \cite{ref6}

\begin{figure}
\centering\includegraphics[width=\linewidth, clip]{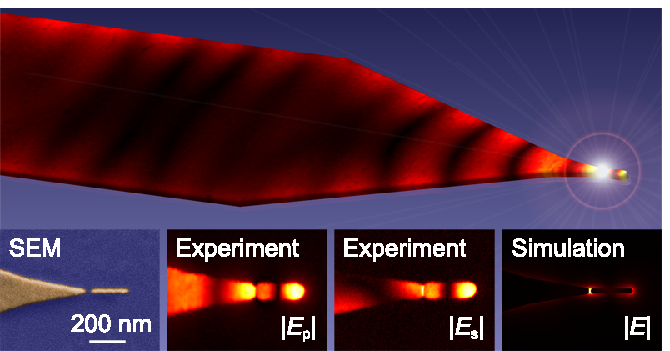}
\end{figure}

\begin{figure*}
\centering\includegraphics{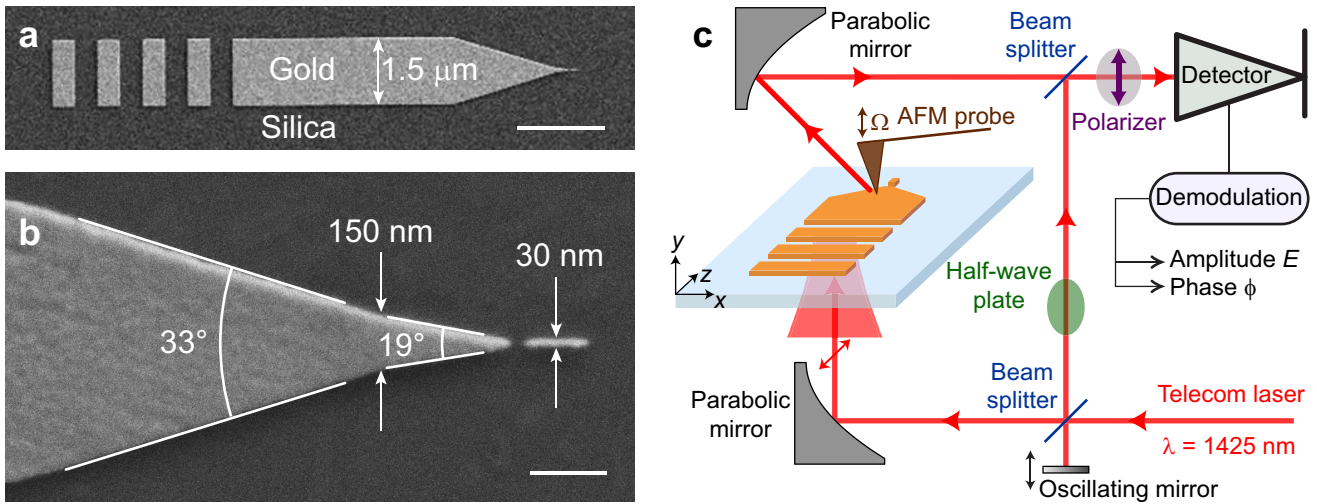}
\caption{(a) SEM image of the fabricated a-TSNF. (b) Zoomed-in tapered part of the structure, showing a two-section taper and an antenna. White bars correspond to 2~$\mu{\rm m}$ in (a) and 200 nm in (b). (c) Schematic layout of the background-free amplitude- and phase-resolved scattering-type SNOM.}
\label{fig1}
\end{figure*}

Nanofocusing using SPP modes was first theoretically described for planar tapered metal-insulator-metal (MIM) and insulator-metal-insulator (IMI) geometries\cite{ref7} and for tapered nanorods, \cite{ref8,ref9} instigating experiments with various configurations demonstrating this fascinating phenomenon (see review by Gramotnev and Bozhevolnyi\cite{ref6} and references therein).  As far as on-chip nanofocusing is concerned, the experimental demonstrations using various SPP configurations and material platforms have been reported. \cite{ref10,ref11,ref12,ref13,ref14,ref15,ref16,ref17,ref18} Arguably the most impressive results were obtained with a MIM waveguide followed by a three-dimensional linear taper that facilitated the nanofocusing of infrared radiation, producing the intensity enhancement of 400 within a subwavelength $14\times 80\ {\rm nm}^2$ area. \cite{ref12} It should however be noted that the considered structure requires the use of rather sophisticated fabrication techniques as well as poses considerable difficulties in accessing the enhanced field, which is mainly distributed within the insulator between the metal stripes. Other configurations (which are less demanding in fabrication but also designed for considerably longer wavelengths) include the nanofocusing using tapered transmission lines, where the moderate intensity enhancement of $\sim 17$ was demonstrated within a hot spot of $\sim 60$ nm diameter, which is equivalent to $\lambda/150$, where $\lambda$ is the free space wavelength. \cite{ref17,ref18} Finally, the nanofocusing has been demonstrated with even simpler (IMI-based) configurations, such as SPP stripe waveguides, in which linearly tapered stripes were used to concentrate the electromagnetic energy carried by short-range SPP modes\cite{ref5} far beyond the diffraction limit with a relatively large efficiency of $\sim50\%$.\cite{ref14,ref15,ref16}

In this paper, we propose, analyze, and experimentally demonstrate efficient on-chip nanofocusing at telecom wavelengths using a tapered-stripe nanofocuser (TSNF) coupled (in the end-fire excitation geometry) to an impedance-matched nanowire antenna across a gap. The nanofocusing effect in the TSNF is thereby combined with and enhanced by the resonance in the nanoantenna, which can be considered as an SPP nanowire analogue of the Fabry-P\'erot interferometer, allowing us to further significantly boost the local intensity enhancement up to about 12,000, with the enhanced field being evenly distributed over the gap volume ($30\times 30\times 10\ {\rm nm}^3$). The taper shape is optimized numerically by maximizing its intensity enhancement and transmittance. The latter is additionally measured experimentally using the amplitude- and phase-resolved scanning near-field optical microscopy (SNOM). Finally, the experimentally studied antenna-coupled TSNF (a-TSNF) provides the evidence of resonant impedance-matched excitation of the nanoantenna by featuring a considerable local-field enhancement (FE) and substantial resonant decrease of back reflection from the tip of the taper, which is similar to the reflection suppression from a Fabry-P\'erot interferometer under the resonance conditions.

The design of our nanofocusing structures (Figure~\ref{fig1}a,b) is based on a 30-nm-thick gold stripe waveguide. The sample was fabricated using a combination of the electron-beam lithography and lift-off technique on a silica substrate (see Supporting Information for details). The width of the widest part of the taper ($w_{\rm in} = 1.5\ \mu{\rm m}$) was chosen to be close to the free-space wavelength of the illumination ($\lambda = 1425$~nm). The width of the narrowest part of the taper ($w_{\rm out} = 30$~nm) is equal to the metal thickness in order to avoid fabrication difficulties. The metal thickness was chosen as a trade-off between the field localization and Ohmic losses. It should be noted that, for the chosen range of the metal stripe width, only one guided mode is supported, which is convenient for observation of a nanofocusing effect. The SPP stripe mode was excited by the laser beam normally incident on the grating (Figure~\ref{fig1}a) designed to have the period equal to the wavelength of the guided plasmonic mode $\lambda_{\rm p} \approx 1.0\ \mu{\rm m}$. In order to suppress the effects of any leaky or radiating modes, excited by diffraction at the grating, a 5-$\mu{\rm m}$-long stripe was placed between the grating and the taper entrance.

\begin{figure*}
\centering\includegraphics{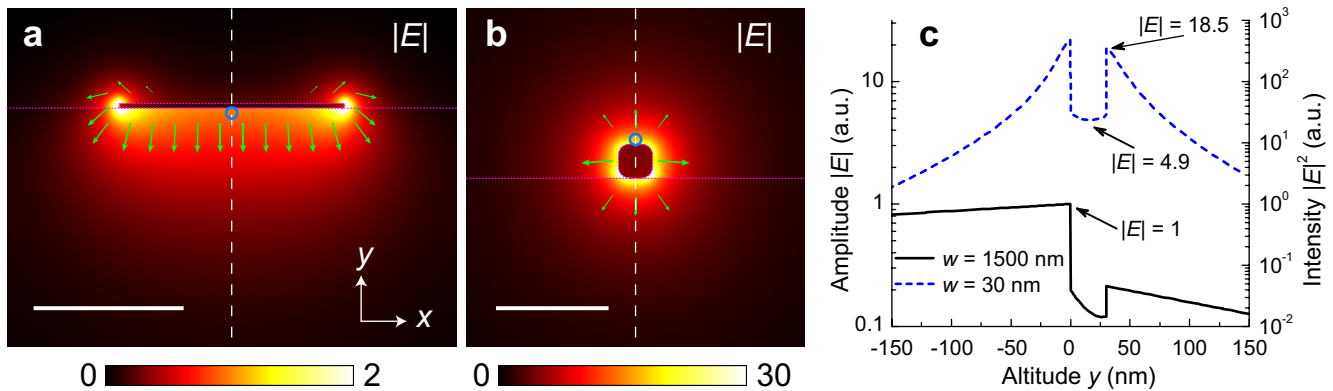}
\caption{Electric-field amplitude distributions in the guided plasmonic mode at the stripe widths of: (a) $1.5~\mu{\rm m}$; and (b)~30~nm. White bars correspond to $1~ \mu{\rm m}$ in (a) and 100 nm in (b). The $E$-field vectors are shown by the green arrows. Dotted magenta lines represent the surface of the silica substrate and gold stripe. The blue circles show the probe points. Field in (a) is normalized to the electric-field amplitude at the probe point. Field in (b) is normalized to give the same total power flow as in (a). (c) Amplitude and intensity profiles, taken along the white dashed lines in (a) and (b).}
\label{fig2}
\end{figure*}

The structures were experimentally investigated using a scattering-type SNOM, based on an atomic force microscope (AFM). We used our SNOM in the transmission mode\cite{ref19,ref20} with the sample illuminated normally from below (Figure~\ref{fig1}c). SNOM with a platinum-coated AFM Si probe was operating in a tapping mode at frequency $\Omega \approx  250$~kHz, while the sample was moving during the scan. The lower parabolic mirror was moving synchronously with the sample in the $xz$-plane in order to maintain the excitation alignment during the scan. The signal, scattered by the probe, was detected and demodulated at the third harmonic ($3\Omega$) to filter the near-field contribution from the background. Additionally, our SNOM uses interferometric pseudoheterodyne detection, \cite{ref21} which allows imaging of both the amplitude and the phase of the near-field. Finally, it should be mentioned that because of the AFM probe shape, elongated along the out-of-plane $y$-axis, the measured optical signal corresponds mostly to the $E_y$ component of the near-field. \cite{ref20,ref22,ref23}

It is known that there are certain constraints on the taper shape for adiabatic nanofocusing. \cite{ref6,ref9} However, the Ohmic losses of plasmonic modes increase with mode compression. Therefore, in order to maximize the efficiency of the taper, one should balance between Ohmic and scattering losses due to non-adiabaticity. We performed numerical optimization of the taper shape, fixing the input and output widths at $w_{\rm in} = 1.5\ \mu{\rm m}$ and $w_{\rm out} = 30$~nm, correspondingly. First, the mode analysis for the two stripe widths $w_{\rm in}$ and $w_{\rm out}$ was performed, using the finite element method (Figure~\ref{fig2}). The mode profile in the wide stripe (Figure~\ref{fig2}a) demonstrates that the electric field is mostly concentrated in the substrate below the stripe, while on the air side it is noticeable only near the edges. For the narrow stripe (called nanowire below), the mode profile in Figure~\ref{fig2}b demonstrates nearly uniform distribution of the electric field around the waveguide with “radial” polarization.\cite{ref3} Thus the nanofocusing of plasmons along the tapered stripe is topologically the same as nanofocusing along three-dimensional metal cones. \cite{ref16} In order to quantify the nanofocusing effect correctly without influence of the lightning-rod effect near the sharp edges, \cite{ref24} we set the probe point (blue circle in Figure~\ref{fig2}a) just below the wide stripe in the silica substrate. For the nanowire (Figure~\ref{fig2}b), the probe point was placed just above the nanowire in the air, where it is accessible for applications. From the middle cross-sections (Figure~\ref{fig2}c) one can see that, for the same total transmitted power, the electric-field amplitude in the nanowire mode (the dashed curve in Figure~\ref{fig2}c) is 18.5 times larger than the field at the wide section of the stripe (the solid curve in Figure~\ref{fig2}c), meaning the intensity enhancement of $\sim 18.5^2 \approx  340$ (under the assumption of zero losses). Interestingly, this intensity enhancement was achieved upon just the 50-fold reduction of the taper width from 1500 nm to 30 nm. The mismatch between the reduction of the taper width and intensity enhancement ($340 >> 50$) can be explained by the two-dimensional mode compression in the waveguide tapered in one dimension (compare Figures~\ref{fig2}a and \ref{fig2}b).

A numerical optimization of the taper shape by maximizing its intensity enhancement and transmittance (see Supporting Information, Figures S1-3) resulted in a combined two-section taper (Figure~\ref{fig1}a,b), whose transmittance was $\sim 41\%$. Here, transmittance is defined as the ratio of the energy flow along the $z$-axis in the guided SPP mode at the taper width of 30 nm to that at the beginning of the taper (Figure~\ref{fig1}a,b). Interestingly, the losses of $59\%$ were distributed almost equally between Ohmic losses in the metal ($27\%$) and scattering into the substrate (31\%), while only $\sim 1\%$ of the energy was reflected back. Thus, the expected intensity enhancement of a lossy TSNF is $\sim 340\times 0.41 \approx 140$. Note that the optimum taper shape (including the angles of each section) depends on the stripe thickness, free-space wavelength, etc. Therefore, thorough numerical simulation should be performed for each particular case to find its own optimum shape.

\begin{figure*}
\centering\includegraphics{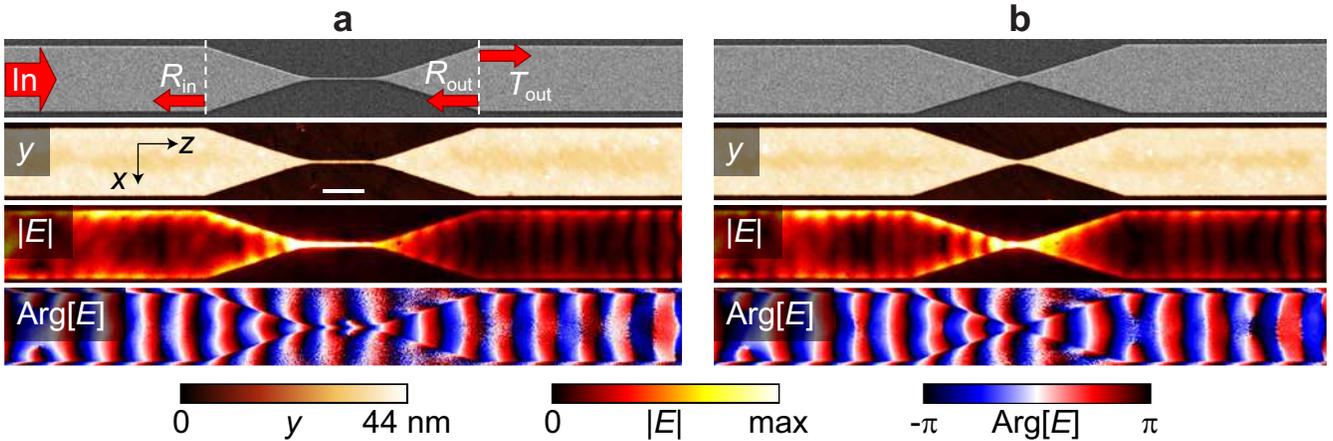}
\caption{(a, b) From top to bottom: SEM and SNOM images of topography y, near-field amplitude $|E|$ and phase Arg[$E$] of the in-coupling and out-coupling TSNFs connected by nanowires of length (a) 1.5~$\mu{\rm m}$ and (b) 0~$\mu{\rm m}$, respectively. The horizontal scale bar in (a) corresponds to 1~$\mu{\rm m}$ and applies to all images. The input SPP stripe mode, transmission and reflections waves are illustrated in SEM image, (a).}
\label{fig3}
\end{figure*}

To experimentally characterize the TSNF transmittance, we fabricated structures with two TSNFs, connected with a nanowire (Figure~\ref{fig3}). The length of the middle nanowire was 1.5~$\mu \rm m$ in one case (Figure~\ref{fig3}a) and zero in the other (Figure~\ref{fig3}b). As can be seen from the near-field amplitude images, transmission through the two TSNFs connected by the nanowires is relatively high. The complex near-field pattern of each section with the wide stripe is due to interference of the guided, leaky and radiation modes. Similarly to our previous work, \cite{ref19} the interference pattern was carefully analyzed and decomposed into individual modes by numerical fitting procedure (see Supporting Information, Figures S13-15). At the same time, the domination of the main guided mode (direct and reflected) is demonstrated by the bright fringes near the edges of the wide stripe, which agrees well with its mode profile (Figure~\ref{fig2}a). Assuming the in-coupling and out-coupling transmittances of the TSNF are the same (supported by numerical simulations – see Supporting Information, Figure~S3), the transmittance for a single TSNF is $T_{\rm exp} \sim 56\%$, which is in good agreement with the numerical simulations.

Once an SPP stripe mode has been nanofocused into a strongly confined guided mode of a nanowire, a few additional options can be used to further enhance the local electric field. Simple back-reflection of the focused SPP mode can cause additional $\sim$ 2-fold FE ($\sim$ 4-fold for the intensity enhancement) because of constructive interference. \cite{ref12} Another option is to introduce a small gap in the metal waveguide. In this case, the longitudinal component of the electric field inside metal can be enhanced in the gap (filled with dielectric) because of the boundary conditions by a factor equal to the ratio of the dielectric constants of the metal and dielectric. Considering a gold waveguide ($\varepsilon_{\rm Au} = -81.6 +8.7i$ at $\lambda$ = 1425~nm\cite{ref25}) with an air gap ($\varepsilon_{\rm air} = 1$), one would expect $\sim$80-fold increase of the longitudinal electric field component inside the gap. Since for a $30\times 30\ {\rm nm}^2$ nanowire the longitudinal component of the electric field inside the metal is $\sim$3.8 times smaller than the total electric field in the air on the nanowire surface (Figure~\ref{fig2}b), the gap-induced FE is $80/3.8 \approx  21$ ($\sim$440 for intensity enhancement). However, our numerical simulations (see Supporting Information, Figure S4) show that, to achieve these effects, the gap should be sufficiently small – for example, a 10 nm gap results in $\sim$60\% reflectance and FE of only $\sim$5.8.

\begin{figure}[htb]
\centering\includegraphics{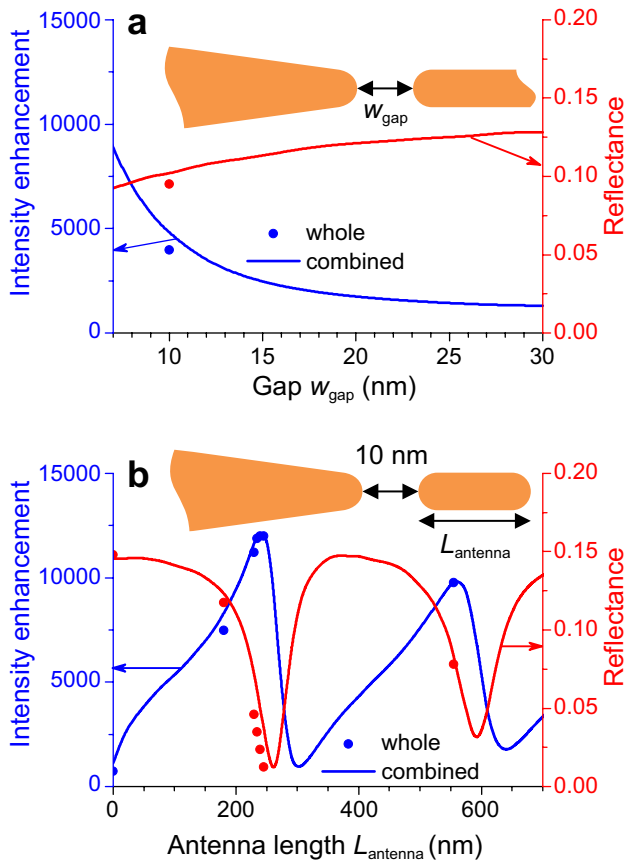}
\caption{Plots of the intensity enhancement inside the gap (blue) and reflectance (red) of (a) g TSNF as a function of the gap width $w_{\rm gap}$ and (b) a-TSNF as a function of antenna length $L_{\rm antenna}$ at the fixed gap width $w_{\rm gap}$ = 10 nm. Insets show simulated configurations. Both figures are calculated at $\lambda$ = 1425 nm. Dots represent simulations of the whole structures, while solid lines represent combination of separate simulations of the TSNF and the gap-induced (antenna-coupled) nanowire.}
\label{fig4}
\end{figure}

The gap-induced FE can be combined with nanofocusing by introducing a gap at the tip of the taper, forming a gapped TSNF (g-TSNF, Fig.~\ref{fig4}a). Its intensity enhancement (solid blue line) can be calculated as a product of TSNF intensity enhancement and a square of the gap-induced FE, which agrees well with the simulations of the whole structure (blue dots). Thus, a 10 nm gap results in g-TSNF intensity enhancement of $\sim 140\times 5.8^2 \approx  4700$. Similarly, the reflectance (red) can be estimated as a product of the gap-induced reflectance in nanowire and a square of TSNF transmittance.

Finally, even greater FE could be achieved when using a dipole nanoantenna coupled to the tip of a TSNF across a narrow nanogap that plays the role of a semi-transparent mirror in a conventional Fabry-P\'erot interferometer. In this case of an antenna-coupled TSNF (a-TSNF), the nanofocusing effect is combined with the antenna resonance, producing particularly strong local fields in the gap between the tip of the TSNF and the nanoantenna. This is also the geometry of the end-fire excitation of a dipole nanoantenna by means of a nanofocusing structure in the on-chip configuration. Expectedly, the maximum of the field inside the gap approximately corresponds to the minimum of the back reflection (Figure~\ref{fig4}b) – small relative extremum displacements can be explained by the Ohmic and radiative losses. \cite{ref26,ref27} Note, however, that the indicated analogy with the Fabry-P\'erot interferometer is not complete. For example, one of the significant physical differences is the strong enhancement of the local field in the nanogap compared to the rest of the structure, which is not typical for a conventional Fabry-P\'erot resonator with mirrors and bulk waves.

The resonant excitation of a 240-nm-long nanoantenna leads to antenna-induced FE of $\sim$9.3 (see Supporting Information, Figure S5), which corresponds to the overall a-TSNF intensity enhancement of $\sim 140\times 9.3^ 2 \approx  12000$. This is $\sim$10 times larger than for the terminated TSNF (i.e., at zero antenna length in Figure~\ref{fig4}b) and $\sim$2.5 times larger than for the g-TSNF (i.e., with an infinitely long nanoantenna). In addition to the described particularly strong local field enhancement in the gap, this local field is highly uniform within the volume of the gap (see Supporting Information, Figure S6), which is the consequence of the quasi-static approximation, where the nanogap has similarity with a parallel plate capacitor with nearly uniform field inside. This could be regarded as a practical advantage of the considered a-TSNF over other designs relying on the lightning-rod effect at sharp corners.

\begin{figure*}
\centering\includegraphics{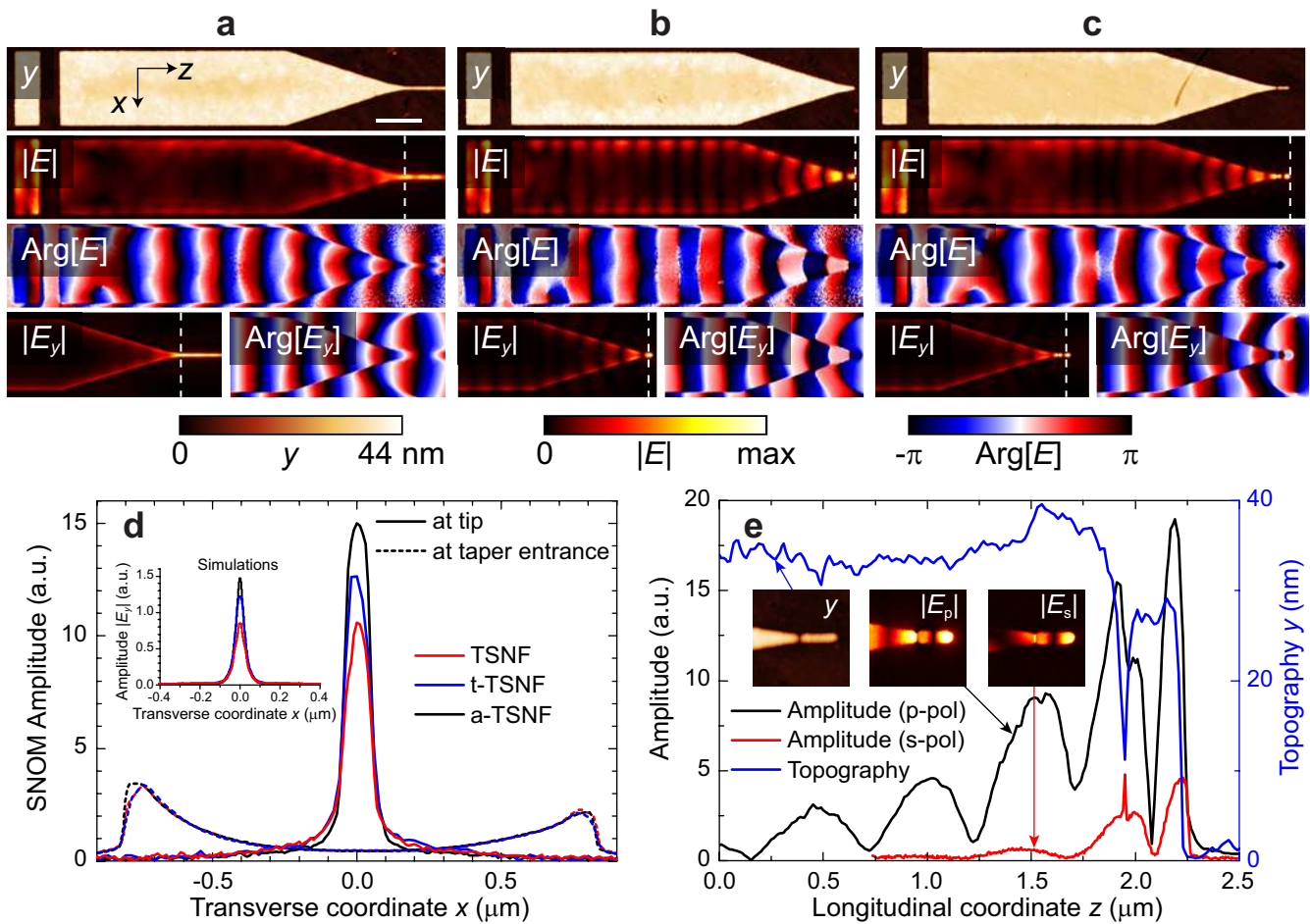}
\caption{(a-c) From top to bottom: SNOM images of topography $y$, measured SNOM near-field (amplitude $|E|$ and phase Arg[$E$]) and simulated field $E_y$ (amplitude $|E_y|$ and phase Arg[$E_y$], taken at the altitude $y$ = 45 nm above the silica surface) for (a) TSNF, (b) t-TSNF and (c) a-TSNF. The 10 nm gap and 235 nm antenna length are used for the simulations and in the design for the fabrication. The scale bar in (a) corresponds to 1~$\mu{\rm m}$ and is the same for the other images (a-c). The near-field amplitudes were normalized to the amplitude of the direct-propagating plasmonic stripe mode at the entrance to the taper. (d) SNOM amplitude profiles (the solid curves) taken along the white dashed lines in (a-c), corresponding to the maximum field amplitudes of TSNF (red), t-TSNF (blue), and a-TSNF (black) structures; dashed curves give the x-profiles of the plasmonic stripe modes at the entrance to the taper for the three structures in (a-c). The inset shows the same line profiles for numerical simulations. (e) Topography (blue) and near-field amplitude profiles taken along the axis of the a-TSNF for the two different detection polarizations: s polarization (red), and p-polarization (black). The insets show the corresponding SNOM images (panel’s lateral size being 750 nm). All measurements are done at $\lambda$ = 1425 nm.}
\label{fig5}
\end{figure*}

We fabricated and experimentally investigated the TSNF (Figure~\ref{fig5}a), the terminated TSNF (t TSNF, Figure~\ref{fig5}b), and the a-TSNF (Figure~\ref{fig5}c) at $\lambda$ = 1425 nm. SNOM measurements were compared to numerical simulations by taking a profile of the out-of-plane $E_y$ field component at the altitude of $y$ = 45 nm above the silica surface (i.e., 15 nm above the top gold surface). The $E_y$ field component is chosen due to the polarization selectivity of the SNOM, dictated by the elongation and corresponding polarizability of the AFM tip along the out-of-plane $y$-axis. The final non-zero altitude of the profile section is chosen in order to take into account the complicated transfer function of the SNOM (see Supporting Information). \cite{ref28} As in the previous section, the measured complex near-field pattern of each section with the wide stripe was decomposed into individual guided and leaky modes by the numerical fitting procedure (see Supporting Information, Figure S12). Then the amplitude of the direct-propagating guided stripe mode was used for normalization of the near-field amplitudes. Their x-profiles at the beginning of the taper (the dotted curves in Figure~\ref{fig5}d) show close similarity, which reflects the robustness of both the fabrication and fitting procedures.  The profiles with the maximum near-field amplitudes for all three structures (the solid curves in Figure~\ref{fig5}d), taken along the white dashed lines in Figure~\ref{fig5}a-c, demonstrate substantial FE compared to the field of the wide stripe mode. At the same time, the taper-gap-antenna structure demonstrates not only the excitation of the nanoantenna (Figure~\ref{fig5}c), but also the strongest observed field enhancement among the three structures (Figure~\ref{fig5}d). This observed relative FE (e.g., 1.2 for the a-TSNF, compared to the t-TSNF) should not be confused to the total FE, expected from the simulations (e.g., $10^{(1/2)} \approx  3.2$ for the a-TSNF, relative to the t-TSNF), since SNOM measurements do not provide the field inside the gap, where the FE is the strongest. The appropriate line profiles, extracted from the simulated field distributions (inset in Figure~\ref{fig5}d), agree well with the experimental results. Further evidence of resonant antenna excitation was the reduction of back reflection. As can be observed, the t-TSNF produces the most pronounced fringes in the near-field amplitude map (Figure~\ref{fig5}b), indicating the largest back-reflection. The numerical fitting procedure yielded a substantially small reflectance of 4\% for a-TSNF compared to $\sim$25\% for t-TSNF.

The expected strong FE inside the gap between the nanoantenna and the tip of the TSNF is not observed in the near-field profiles (Figure~\ref{fig5}c) because of the polarization selectivity of the SNOM, dictated by the elongation and corresponding polarizability of the AFM tip along the out-of-plane $y$-axis. In order to overcome this restriction, we set a polarizer in front of the detector (Figure~\ref{fig1}c) in order to suppress the scattered $E_y$ component and access the in-plane near-field component $E_z$ (the detected signal is denoted as s-polarization). The detected signal with the orthogonal p-polarization, achieved by 90-degrees rotation of the polarizer, corresponds to the maximum contribution of $E_y$ near-field. The recorded near-field maps of the nanoantenna in a TSNF (Figure~\ref{fig5}e) clearly show the sharp peak in the $|E_{\rm s}|$ profile at $z = 1.9~\mu{\rm m}$, exactly corresponding to the gap, that provides further evidence of the FE inside the gap. The comparison of the measured $E_{\rm p}$ and $E_{\rm s}$ profiles with the simulated field distribution is given in Supporting Information, Figure S17.

In summary, we have demonstrated a simple and robust planar configuration, which is amenable for one-step lithography fabrication, for efficient on-chip nanofocusing of radiation carried by stripe SPP modes. The considered structure, consisting of a tapered metal stripe coupled across a gap to a stripe nanoantenna, combines the benefits of near-adiabatic SPP nanofocusing in an optimized tapered structure (characterized by weak scattering losses) with the efficient resonant end-fire excitation of a dipole antenna mode, resulting in further boosting the local FE in the gap. Because of the unique combination of the nanoantenna resonance and optimized SPP nanofocusing (with $\sim$41\% energy transmission through the stripe taper and the corresponding intensity enhancement of $\sim$140), the total intensity enhancement in the gap was deduced to be $\sim$12000 with a nearly uniform field over the gap volume (of $30\times 30\times 10~{\rm nm}^3$). Compared to the excitation of the dimer antenna of similar size under the normal illumination (see Supporting Information, Figure S8), the a-TSNF shows $\sim$1.3 times larger intensity enhancement and about 3 times larger $|E|^2$ inside the gap, normalized to the same total incident power. Supplemented with compactness and background-free hotspot excitation due to the on-chip realization, the a-TSNF design looks superior to the common dimer antenna design. 

We would like to emphasize that our concept of a tapered structure (enabling radiation nanofocusing) being coupled (across a gap) to a resonant nanoantenna for further boosting the FE is rather general, and can be applied to many different SPP-based waveguide configurations, such as MIM waveguides, \cite{ref12} slot waveguides, \cite{ref10} transmission lines, \cite{ref17,ref18} hybrid plasmonic-photonic waveguides, \cite{ref11} etc. Additionally, the proposed end-fire coupling can be used for excitation of any nanoantenna, including those supporting dark modes. The latter is illustrated with our simulations shown in Figure~\ref{fig4}b, where the second resonance ($L_{\rm antenna} \approx  560$~nm) corresponds to the second-order antenna mode, which is a dark mode for the far-field excitation. Finally, it should be borne in mind that both nanofocusing and antenna-coupling approaches are not limited to the visible and infrared wavelength ranges, but can also be applied in the THz range, \cite{ref29, ref30, ref31} implying the possibility of making use of their combination considered in this work for efficiently concentrating and strongly enhancing the THz radiation in subwavelength-sized hot spots.

The considered combination of a tapered SPP waveguide and a gap-coupled nanoantenna opens up new avenues in the development of integrated nano-optical devices and circuits with a variety of useful on-chip functionalities, in which nanofocusing tapers could serve as efficient inputs/outputs for SPP signals that are to be used for further processing or detection. For example, a nanowire antenna (or a chain of such antennae) coupled to the input and output stripe tapers could be used for spectrally selective transmission (see Supporting Information, Figure S7), thus giving a possibility of combining SPP nanofocusing with wavelength filtering functionalities. In the case of two antennae (three gaps) we observe two local spectral maxima of transmittance corresponding to two different electric field distributions over the three gaps (see Supporting Information). This offers a possibility of adjusting the spatial position of subwavelength-sized hot spots, which could be useful for the realization of efficient spatio-temporal control on the nanoscale. A combination of strong local FE that can be obtained with our configuration and nonlinearities of the dielectric in the gap(s) or intrinsic nonlinearities of metal (gold) could open further opportunities for the development of new nanoscale non-linear components. Overall, bearing in mind the relative simplicity in fabrication and aforementioned features of the proposed structures, we believe that the unique combination of efficient nanofocusing and nanoantenna resonant excitation realized in our experiments promises a variety of useful on-chip functionalities within sensing, nonlinear spectroscopy and signal processing.

\textbf{Acknowledgments}

V.A.Z., I.P.R., V.S.V., and S.I.B. acknowledge financial support from the Danish Council for Independent Research (the FTP project ANAP, Contract No. 09-072949) and from the European Research Council, Grant No. 341054 (PLAQNAP). A.A. acknowledges financial support from the Danish Council for Independent Research via the GraTer project (Contract No. 0602-02135B).

\textbf{Supporting Information Available}

Details on the fabrication procedure, numerical simulations and optimization, processing of the experimental data. This material is available free of charge via the Internet at \url{http://pubs.acs.org}.

\end{document}